\documentclass[a4paper,fleqn,usenatbib]{mnras}

\usepackage{newtxtext,newtxmath}
\usepackage[T1]{fontenc}
\usepackage{ae,aecompl}
\usepackage{graphicx}
\usepackage{amsmath}
\usepackage{amssymb}
\usepackage{subfig}

\newcommand{\Msun}{\ensuremath{\,{M}_\odot}}                            
\newcommand{\Rsun}{\ensuremath{\,{R}_\odot}}                            
\newcommand{\Mjup}{\ensuremath{\,{M}_{\rm Jup}}}                        
\newcommand{\Rjup}{\ensuremath{\,{R}_{\rm Jup}}}                        
\newcommand{\Teq}{\ensuremath{T_{\rm eq}^{\,\prime}}}                   
\newcommand{\safronov}{\ensuremath{\Theta}}                             
\newcommand{\mss}{\,m\,s$^{-2}$}                                        
\newcommand{\pjup}{\ensuremath{\,\rho_{\rm Jup}}}                       
\newcommand{\psun}{\ensuremath{\,\rho_\odot}}                           

\title[Physical properties of WASP-74\,b]{Physical properties and transmission spectrum of the WASP-74 planetary system from multi-band photometry\thanks{Based on data collected by MiNDSTEp with the Danish 1.54\,m telescope at the ESO La Silla Observatory.}}

\author[L. Mancini et al.]
{\parbox{\textwidth}{L. Mancini$^{1,\,2,\,3,\,4}$\thanks{E-mail: \href{lmancini@roma2.infn.it}{lmancini@roma2.infn.it}},
J.\ Southworth$^{5}$,
P.\ Molli\`{e}re$^{6}$,
J.\ Tregloan-Reed$^{7}$,
I.\,G. Juvan$^{8,\,9,\,10}$,
G.\ Chen$^{11,\,12}$,
P.\ Sarkis$^{2}$,
I.\ Bruni$^{13}$,
S.\ Ciceri$^{14}$,
M.\ I.\ Andersen$^{15,\,16}$,
V.\ Bozza\,$^{17,\,18}$,
D.\ M.\ Bramich$^{19}$,
M.\ Burgdorf$^{20}$,
G.\ D'Ago$^{21}$,
M.\ Dominik$^{22}$,
D.\ F.\ Evans$^{5}$,
R.\ Figuera Jaimes$^{22,\,23,\,24}$,                 
L.\ Fossati$^{8}$,
Th.\ Henning$^{2}$,
T.\ C.\ Hinse$^{25,\,26}$,
M.\ Hundertmark$^{27}$,
U.\ G.\ J{\o}rgensen$^{16}$,
E.\ Kerins$^{28}$,
H.\ Korhonen$^{15,\,16}$,
M.\ K\"uffmeier$^{29}$,
P.\ Longa$^{7}$,
N.\ Peixinho$^{30}$,
A.\ Popovas$^{16}$,
M.\ Rabus$^{21,\,2,\,31,\,32}$,
S.\ Rahvar$^{33}$,
J.\ Skottfelt$^{34,\,16}$,
C.\ Snodgrass$^{35}$,
R.\ Tronsgaard$^{36,\,37}$,
Y.\ Wang$^{38,\,11}$,
O.\ Wertz$^{39}$\\
}
\\
Affiliations are listed at the end of the paper.
}

\date{Accepted XXX. Received YYY; in original form ZZZ}

\pubyear{2018}

\begin{document}
\label{firstpage}
\pagerange{\pageref{firstpage}--\pageref{lastpage}}
\maketitle

\vspace{10cm}

\begin{abstract}
We present broad-band photometry of eleven planetary transits of the hot Jupiter WASP-74\,b, using three medium-class telescopes and employing the telescope-defocussing technique. Most of the transits were monitored through $I$ filters and one was simultaneously observed in five optical ($U, g^{\prime}, r^{\prime}, i^{\prime}, z^{\prime}$) and three near infrared ($J$, $H$, $K$) passbands, for a total of 18 light curves. We also obtained new high-resolution spectra of the host star. We used these new data to review the orbital and physical properties of the WASP-74 planetary system. We were able to better constrain the main system characteristics, measuring smaller radius and mass for both the hot Jupiter and its host star than previously reported in the literature. Joining our optical data with those taken with the HST in the near infrared, we built up an observational transmission spectrum of the planet, which suggests the presence of strong optical absorbers, as TiO and VO gases, in its atmosphere.

\end{abstract}

\begin{keywords}
stars: planetary systems -- stars: fundamental parameters -- stars: individual: WASP-74 -- techniques: photometric
\end{keywords}


\section{Introduction}
\label{sect_01}

Transiting hot Jupiters form a class of exoplanets that has been extensively studied, especially those orbiting bright ($V<14$\,mag) stars. Because of their large size ($R_{\rm p}\approx 1\,R_{\rm Jup}$) and short orbital periods ($P_{\rm orb} < 10$\,d), hot Jupiters are more easily detected in transit, even by ground-based surveys with small telescopes or telephoto lenses. Indeed, their transits are very frequent and their large sizes imply light curves with transit depths at the percent level (see e.g. \citealp{cameron:2016}). Moreover, the possibility to measure both their mass and radius with exquisite precision makes them excellent objects for a number of theoretical and observational investigations (e.g.\ formation, evolution, dynamics, star-planet and planet-planet interaction, orbital migration, internal structure, etc.). The great effort made by astronomers to find these exoplanets is still continuing (we currently know more than 350 well-studied transiting hot Jupiters\footnote{Data taken from the Transiting Extrasolar Planet Catalogue (TEPCat), which is available at http://www.astro.keele.ac.uk/jkt/tepcat/ \citep{southworth:2011}.}) and is revealing an incredible diversity in terms of physical characteristics and planetary architectures (see e.g. \citealp{winn:2015}).

Hot Jupiters are also particularly suitable for atmospheric investigation, which is one of the most fascinating perspectives. By monitoring their transit events with large-class telescopes, one can perform the transmission-spectroscopy method, i.e. measure their transit depths at multiple wavelengths, and constrain the chemical composition at the terminator of their atmosphere. It is thus possible to search for the presence of several molecules (e.g.\ H$_2$O, CO, CO$_2$, CH$_4$) at infrared (IR) wavelengths and atoms (e.g.\ Na and K) and other inorganic chemical compounds (TiO and VO) at the optical wavelengths (see, e.g., the recent reviews by \citealp{sing:2018,fortney:2018}).

Even though transmission spectroscopy is by now a common observational technique and is performed both from ground and space, it is still very challenging, because the required ultra-high precision spectroscopic measurements are plagued by systematic effects that are gradually becoming understood and correctable to the required level. Significant progress in this field have been obtained in the last years and the resulting transmission spectra of a small group of hot Jupiters has revealed an unexpected diversity of their atmospheres and different amounts of aerosols at observable pressure levels (e.g. \citealp{sing:2016}). Yet, discrepancies between the results found by different instruments and between ground-based and space observations continue to emerge (e.g. \citealp{sedaghati:2017,parviainen:2018,gibson:2018}), also suggesting a degeneracy of the transmission-spectrum slope with orbital inclination and $R_{\star}/a$ \citep{alexoudi:2018}.

Probing exoplanetary atmospheres with photometric observations represents an alternative approach. The transmission-photometry technique has the clear disadvantage of a lower spectral resolution, but it is generally less affected by telluric contamination and systematics, and can be performed from the ground with smaller-aperture telescopes, which are, generally, more easily accessible than larger facilities. Moreover, multi-band photometry allows us to probe the atmosphere of exoplanets hosted by relatively faint stars ($V\lesssim13$\,mag), which are generally very arduous for spectroscopy.

In the last decade, we have undertaken a large observational program to obtain high-quality transit light curves for many known transiting exoplanets. For this purpose, we have used an array of medium-sized telescopes, located in both hemispheres, and pioneered the defocussing technique and other observational strategies, like the simultaneous two-telescope and multi-band observations. Besides reviewing the orbital and physical parameters of these planetary systems in a homogeneous way, which is the main aim of our project, we have also made use of multi-band imaging cameras to perform transmission photometry for probing planetary atmospheres (e.g.\ \citealp{mancini:2016a,mancini:2016b,southworth:2017,tregloan:2018}).

Here we present a study of the transiting planetary system WASP-74 \citep{hellier:2015}, which hosts an inflated hot Jupiter, WASP-74\,b ($M_{\rm p}\approx 1\,M_{\rm Jup}$; $R_{\rm p}\approx 1.5\,R_{\rm Jup}$; $T_{\rm eq}\approx 1900$\,K), orbiting an F9\,V star every 2.1\,d. WASP-74\,b was included in the sample of 30 hot Jupiters observed with the HST/WFC3 camera by \citet{tsiaras:2018} for measuring transmission spectra between 1.1 and 1.7 $\mu$m. Their results for WASP-74\,b are not conclusive, but indicate high-altitude cloud cover or a water depleted atmosphere. Joining these data with our observations, mostly covering the optical band ($386-976$\,nm), we obtained a transmission spectrum of this exoplanet with much wider spectral coverage.

\section{Observation and data reduction}
\label{sec_2}

\subsection{Photometry}
\label{sec_2.1}
Having declination $\delta \approx -1^{\circ}$, WASP-74 can be observed from both hemispheres. In 2015, we monitored seven (six complete and one partial) transits of WASP-74\,b through a Bessell-$I$ filter with the Danish (DK) 1.54\,m Telescope at the ESO Observatory in La Silla. In the same observing season, another two transits were remotely observed with the Zeiss 1.23\,m telescope at the Calar Alto Observatory: the first was completely covered through a Cousins-$I$ filter, while the second was only partially covered through a Johnson-$B$ filter. 

\begin{figure}
\centering
\includegraphics[width=\columnwidth]{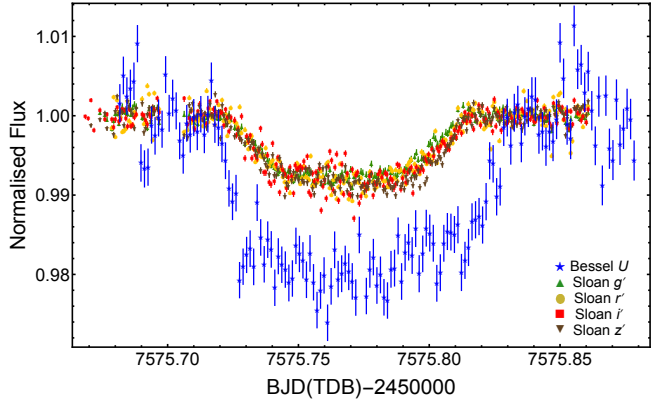}
\caption{Light curves of a transit of WASP-74\,b simultaneously observed with two telescopes: the Danish Telescope ($U$ filter) and the MPG 2.2\,m Telescope ($g^{\prime}$, $r^{\prime}$, $i^{\prime}$, $z^{\prime}$). }
\label{fig:simultransit}
\end{figure}

Another transit event was recorded in 2016 with the MPG 2.2\,m telescope, also located in La Silla, using the GROND instrument to simultaneously observe in four optical (similar to Sloan $g^{\prime}$, $r^{\prime}$, $i^{\prime}$, $z^{\prime}$) and three near-IR (NIR) bands ($J$, $H$, $K$). More information about these telescopes and instruments is available in previous papers (e.g.\ \citealt{mancini:2018}). The same transit was also observed with the Danish telescope using a Bessell-$U$ filter, in order to have a wavelength coverage in the near-ultraviolet as well. The quality of the $U$-band light curve is low, but showed a transit deeper ($2\%$ versus $0.9\%$) and longer (2.8 hr versus 2.4 hr) than those taken with GROND, see Fig~.\ref{fig:simultransit}. A possible interpretation of this fact is planetary mass loss, i.e. WASP-74\,b may be evaporating under the radiation pressure of its host star. In order to confirm this hypothesis, another transit was observed with the Danish Telescope in August 2017, again through a Bessell-$U$ filter, which did not confirm the longer duration of the WASP-74\,b transits in the blue band.

The observations, which we reported above, were all carried out by autoguiding and defocussing the telescopes for increasing the photometric precision.

The reduction of the optical data was performed using the {\sc defot} pipeline \citep{southworth:2014}; master bias and master flat-field frames were produced by median-combining individual calibration images, and used to calibrate the scientific images. A reference image was selected and the target and a set of non-variable comparison stars were identified in it. Pointing variations were measured by a cross-correlation process. Aperture photometry of the target and reference stars was then performed using the {\sc aper} routine\footnote{{\sc aper} is part of the {\sc astrolib} subroutine library distributed by NASA.}, and the aperture radii were chosen to obtain the lowest scatter versus a fitted model. The differential photometry light curves were normalised to zero magnitude by fitting a straight line to the out-of-transit data, but with a second-order polynomial for the $U$-band data. 

The quality of the photometry of both the $U$-band light curves is low because of not optimal (thin clouds) and not stable (continuous variation of seeing concurrently with humidity, temperature and wind speed) weather conditions and a lack of good comparison stars as well. Moreover, $U$-band data are also affected by low throughput of the CCD and $U$-band filter. The resulted light curves are severely affected by systematics, which caused larger transit depth in both cases and, for the first transit, an artificial longer transit duration.

Data reduction of the GROND NIR bands followed the methodology detailed in \citet{mancini:2013} and \citet{chen:2014}. After dark subtraction and flat correction, aperture photometry was performed using the IDL/DAOPHOT package. The choice of aperture and annulus sizes and combination of comparison stars were determined by the light-curve fitting that results in the least scatter. Before the light-curve fitting, variations of star location and full width at half maximum (FWHM) of the point spread function (PSF) were recorded, which will be used to identify the correlated noise. Unfortunately, the pointing jumped three times during the observation. Considering that NIR photometry is strongly location dependent, the data, which were acquired at the time when the pointing was not stabilized, were not used in the subsequent light-curve analysis. This excluded the data points before the gap as seen in the optical light curves and a few other data points at the end of the observation.

All the times of the observations were converted into the BJD(TDB) system using routines from \citep{eastman:2010}. The light curves are plotted in Fig.~\ref{fig:dk:lc}, \ref{fig:grond:lc}, \ref{fig:ca:lc} and \ref{fig:DKU:lc}. The data will be made available from the CDS\footnote{{\tt http://cdsweb.u-strasbg.fr/}}.

\begin{table*}
\centering
\setlength{\tabcolsep}{4pt}
\caption{Details of the transit observations presented in this work. $N_{\rm obs}$ is the number of observations,
$T_{\rm exp}$ is the exposure time, $T_{\rm obs}$ is the observational cadence, and `Moon illum.' is the geocentric
fractional illumination of the Moon at midnight (UT). The aperture sizes are the radii of the software apertures
for the star, inner sky and outer sky, respectively. `Scatter' is the \emph{rms} scatter of the data versus a fitted model. `Bin' refers to if we have binned the data or not. `Compl.' specifies if the transit was completely observed or not.}
\label{tab:obs}
\begin{tabular}{lccccccccccccc} \hline
Telescope & Date of   & Start time & End time  &$N_{\rm obs}$ & $T_{\rm exp}$ & $T_{\rm obs}$ & Filter & Airmass & Moon & Aperture & Scatter & Bin & Compl.\\
               & first obs &    (UT)    &   (UT)    &              & (s)           & (s)           &        &         &illum.& radii (px) & (mmag)  \\
\hline
DK\,1.54\,m & 2015 05 31 & 06:33 & 10:41 & 202 & 60 & 72 & Bessell $I$ & $1.31 \rightarrow 1.13 \rightarrow 1.34$  &  94\%  & 28, 36, 60  & 0.39 & n & y \\ %
DK\,1.54\,m & 2015 06 15 & 06:36 & 10:29 & 190 & 60 & 72 & Bessell $I$ & $1.17 \rightarrow 1.13 \rightarrow 1.56$  & ~~2\%  & 28, 37, 70  & 0.69 & n & y \\  %
DK\,1.54\,m & 2015 06 30 & 05:31 & 09:25 & 484 & 15 & 27 & Bessell $I$ & $1.18 \rightarrow 1.13 \rightarrow 1.53$  &  96\%  & 17, 26, 50  & 0.39 & y & y \\  %
DK\,1.54\,m & 2015 07 15 & 04:07 & 08:38 & 225 & 60 & 72 & Bessell $I$ & $1.23 \rightarrow 1.13 \rightarrow 1.60$  & ~~1\%  & 25, 35, 60  & 0.49 & y & y \\  %
DK\,1.54\,m & 2015 08 29 & 01:52 & 04:22 & 170 & 30 & 42 & Bessell $I$ & $1.16 \rightarrow 1.13 \rightarrow 1.17$  & 100\%  & 23, 32, 60  & 0.66 & n & n \\  %
DK\,1.54\,m & 2015 09 13 & 01:52 & 05:41 & 627 & 16 & 28 & Bessell $I$ & $1.16 \rightarrow 1.13 \rightarrow 2.27$  & ~~0\%  & 20, 30, 60  & 0.59 & y & y \\  %
DK\,1.54\,m & 2015 09 28 & 00:10 & 04:07 & 576 & 10 & 22 & Bessell $I$ & $1.14 \rightarrow 1.13 \rightarrow 1.83$  & 100\%  & 20, 30, 60  & 0.65 & y & y \\ [2pt] %
CA\,1.23\,m & 2015 07 27 & 22:56 & 04:06 & 167 & 100 & 123 & Cousins $I$ & $1.37 \rightarrow 1.15 \rightarrow 2.37$  & 85\%  & 37, 50, 100  & 1.08 & n & y \\ %
CA\,1.23\,m & 2017 09 08 & 19:19 & 01:25 & 153 & 100 & 112 & Johnson $B$ & $1.50 \rightarrow 1.15 \rightarrow 2.45$  & 91\%  & 30, 40, 80  & 1.16 & n & n \\ [2pt] %
%
%
MPG\,2.2\,m & 2016 07 06 & 04:02 & 08:39 &  255 &  30 & 57 & Sloan $g^{\prime}$ & $1.35 \rightarrow 1.13 \rightarrow 1.41$   & ~~4\%  & 55, 70, 200  & 0.69 & n & y  \\ %
MPG\,2.2\,m & 2016 07 06 & 04:02 & 08:39 &  247 &  30 & 57 & Sloan $r^{\prime}$ & $1.35 \rightarrow 1.13 \rightarrow 1.41$   & ~~4\%  & 50, 70, 150  & 1.45 & n & y  \\ %
MPG\,2.2\,m & 2016 07 06 & 04:02 & 08:39 &  262 &  30 & 57 & Sloan $i^{\prime}$ & $1.35 \rightarrow 1.13 \rightarrow 1.41$   & ~~4\%  & 60, 75, 150  & 1.34 & n & y  \\ %
MPG\,2.2\,m & 2016 07 06 & 04:02 & 08:39 &  252 &  30 & 57 & Sloan $z^{\prime}$ & $1.35 \rightarrow 1.13 \rightarrow 1.41$   & ~~4\%  & 60, 80, 200  & 1.09 & n & y  \\ [2pt] %
MPG\,2.2\,m & 2016 07 06 & 04:02 & 08:39 &  421 &  2  & 27 & $J$ & $1.35 \rightarrow 1.13 \rightarrow 1.41$   & ~~4\%  & 18, 18, 25  & 1.33 & n & y  \\ %
MPG\,2.2\,m & 2016 07 06 & 04:02 & 08:39 &  421 &  2  & 27 & $H$ & $1.35 \rightarrow 1.13 \rightarrow 1.41$   & ~~4\%  & 17, 17, 25  & 1.30 & n & y  \\ %
MPG\,2.2\,m & 2016 07 06 & 04:02 & 08:39 &  421 &  2  & 27 & $K$ & $1.35 \rightarrow 1.13 \rightarrow 1.41$   & ~~4\%  & 17, 25, 31  & 2.46 & n & y  \\ %
DK\,1.54\,m & 2016 07 06 & 04:21 & 09:04 & 146 & 100 & 112 & Bessell $U$ & $1.35 \rightarrow 1.13 \rightarrow 1.41$  & ~~4\%  & 09, 19, 25  & 3.39 & n & y \\ %
DK\,1.54\,m & 2017 08 27 & 01:19 & 04:53 & 113 & 100 & 112 & Bessell $U$ & $1.30 \rightarrow 1.14 \rightarrow 1.42$  & 29\%  & 11, 22, 35  & 2.33 & n & y \\ [2pt] %
\hline \end{tabular} \end{table*}

\subsection{Spectroscopy}
\label{sec_2.2}

On June 3, 2017, we acquired three high-resolution spectra using the FEROS \'echelle spectrograph \citep{kaufer:1998}, and used them to refine the stellar parameters of WASP-74. The three spectra were co-added and the stellar atmospheric parameters were estimated using {\sc zaspe} \citep{brahm:2017:zaspe}. {\sc zaspe} determines $T_\mathrm{eff}$, $\log g$, [Fe/H], and $v\sin i$ via least-squares minimisation against a grid of synthetic spectra in the spectral regions most sensitive to changes in the parameters ($5000$\,\AA\ and $6000$\,\AA). We found the following parameters: $T_{\rm eff}=5984 \pm 57$\,K, $\log{g}=4.12 \pm 0.12$\,dex, [Fe/H]$=+0.34 \pm 0.02$\,dex, and $v \sin{i} = 6.03 \pm 0.19$\,km\,s$^{-1}$.

\section{Light curve analysis}
\label{sec_3}

\subsection{Photometric parameters}
\label{sec_3.1}

The optical light curves were modelled using the {\sc jktebop}\footnote{\textsc{jktebop} is written in {\sc fortran77} and is available at: {\tt http://www.astro.keele.ac.uk/jkt/codes/jktebop.html}} code \citep{southworth:2013}. We assumed a circular orbit \citep{hellier:2015} and the following parameters of the light curves were fitted using the Levenberg-Marquardt optimisation algorithm: the sum and ratio of the fractional radii\footnote{The fractional radii are defined as $r_{\mathrm{A}} = R_{\mathrm{A}}/a$ and $r_{\mathrm{b}} = R_{\mathrm{b}}/a$, where $R_{\mathrm{A}}$ and $R_{\mathrm{b}}$ are the true radii of the star and planet, and $a$ is the semi-major axis.} ($r_{\mathrm{A}}+r_{\mathrm{b}}$ and $k=r_{\mathrm{b}}/r_{\mathrm{A}}$), the time of transit midpoint ($T_{0}$), the orbital period and inclination ($P$ and $i$).

\begin{figure*}
\centering
\includegraphics[width=17.0cm]{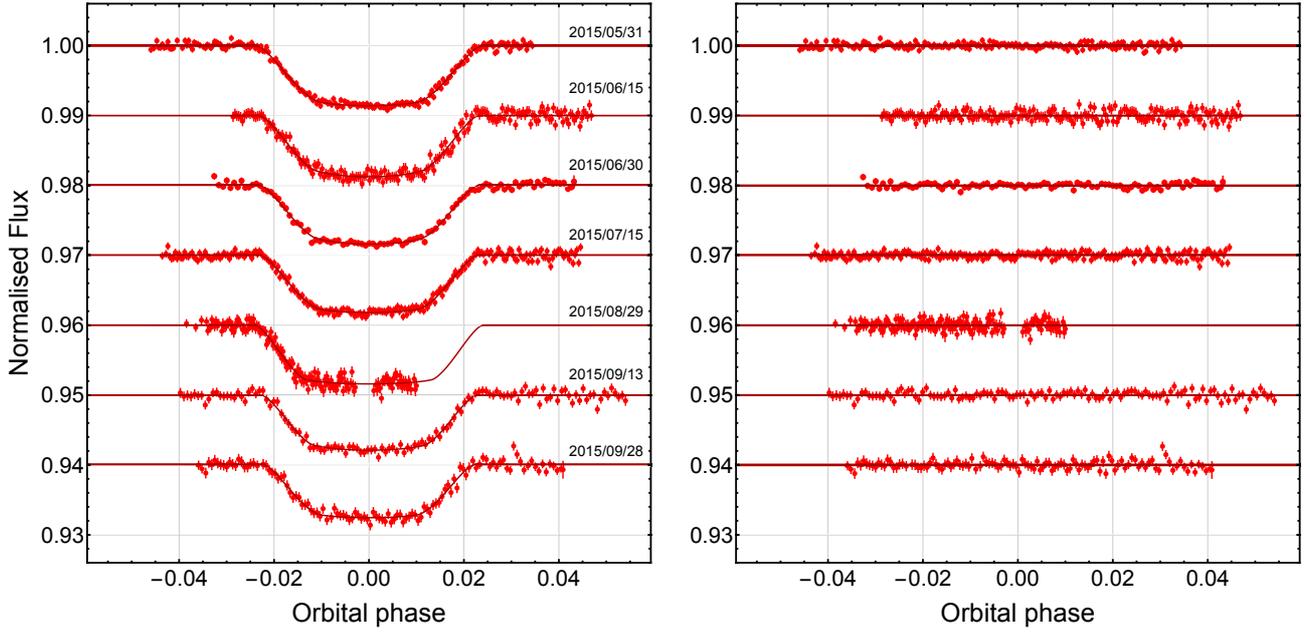}
\caption{{\it Left-hand panel:} light curves of seven transits of WASP-74\,b observed with the Danish 1.54\,m telescope through a Bessell-$I$ filter, shown in date order. They are plotted versus orbital phase and are compared to the best-fitting models.
{\it Right-hand panel:} the residuals of each fit.}
\label{fig:dk:lc}
\end{figure*}

\begin{figure*}
\centering
\includegraphics[width=17.0cm]{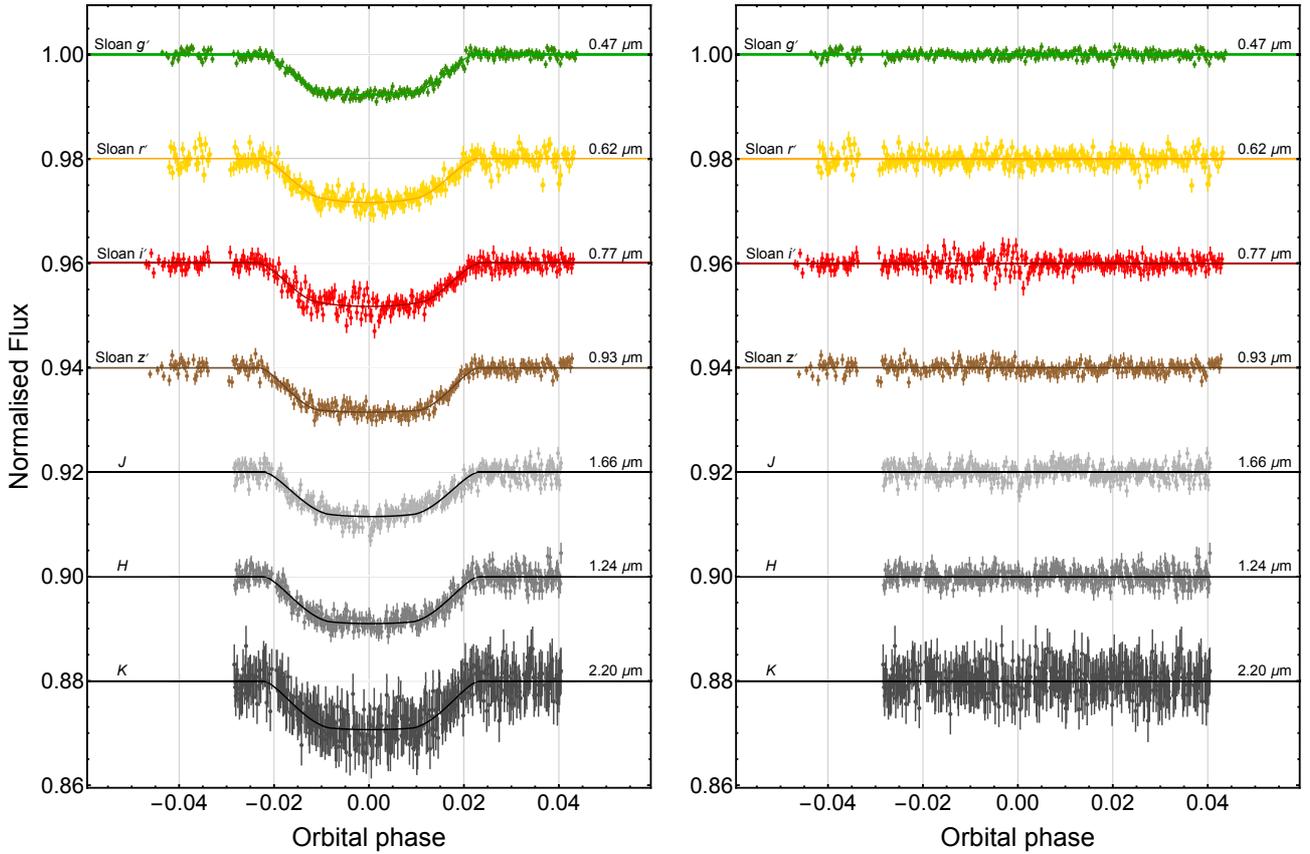}
\caption{{\it Left-hand panel:} simultaneous optical and NIR light curves of one transit event of WASP-74\,b observed with GROND. The best-fitting models are shown as solid lines for each optical data set. The passbands are labelled on the left of the figure and their central wavelengths are given on the right. {\it Right-hand panel:} the residuals of each fit.}
\label{fig:grond:lc}
\end{figure*}

\begin{figure}
\centering
\includegraphics[width=\columnwidth]{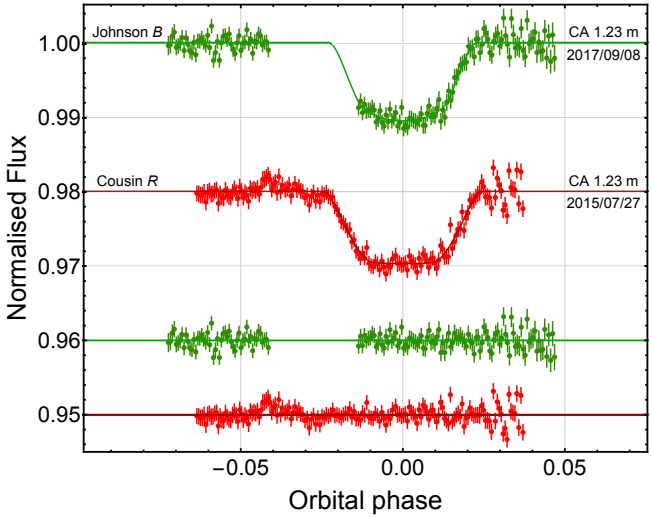}
\caption{Light curves of two transit events of WASP-74\,b observed with the CA 1.23\,m telescope. They are plotted versus orbital phase and are compared to the best-fitting models. The residuals of the fits are shown at the base of the figure. Labels indicate the observation date and the filter that was used for each dataset.}
\label{fig:ca:lc}
\end{figure}

We used a quadratic law to model the limb darkening (LD) of the star, using the LD coefficients provided by \citet{claret:2004}. As in previous papers (e.g.\ \citealt{mancini:2016a}), we fit for the linear coefficient and fixed the non-linear one; this choice is dictated by the fact that the LD coefficients are very strongly correlated (e.g.\ \citealp{southworth:2007}). We included the coefficients of a linear polynomial versus time in the fits in order to take into account the uncertainty in the detrending of the light curves. Finally, we rescaled the uncertainty of the light curve points generated by our reduction pipeline
by multiplying them by the square root of the reduced ${\chi^{2}}$ of the fit. The best fits to the light curves are shown in Fig.\,\ref{fig:dk:lc}, Fig.\,\ref{fig:grond:lc} and Fig.\,\ref{fig:ca:lc}. The parameters of the fits are reported in Table\,\ref{tab:fits}. The uncertainties of the parameters were estimated for each solution from 10\,000 Monte Carlo simulations and through a residual-permutation algorithm. The larger of the two error bars was adopted in each case. The final photometric parameters were calculated as the weighted mean of the results in Table\,\ref{tab:fits}. We also show the values obtained by \citet{hellier:2015} for comparison. 

\begin{figure}
\centering
\includegraphics[width=\columnwidth]{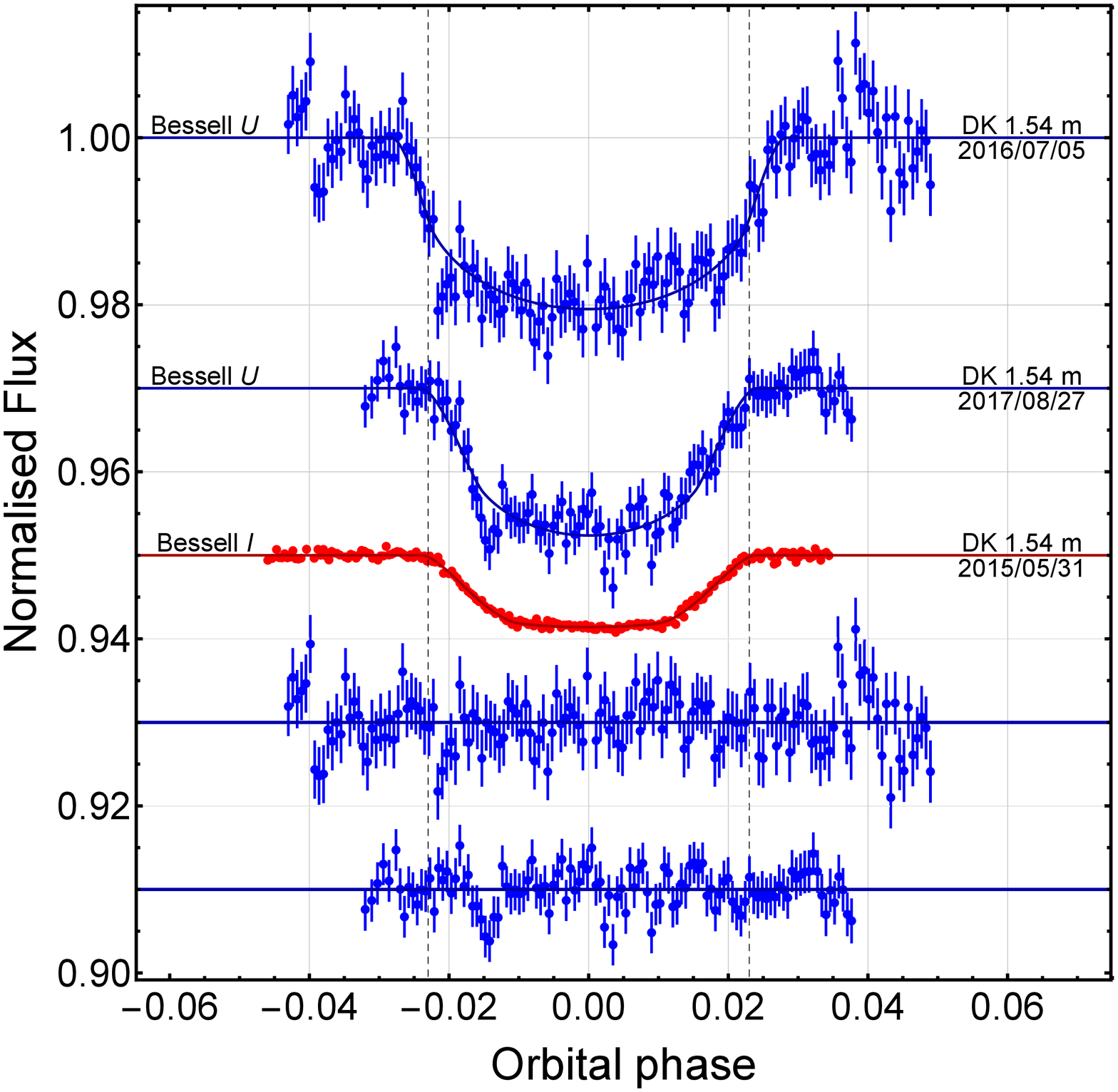}
\caption{Light curves of two transit events of WASP-74\,b observed with the DK 1.54\,m telescope through a Bessell-$U$ filter, shown in date order. They are plotted versus orbital phase and are compared to the best-fitting models. The residuals of the fits are shown at the base of the figure and are preceded by the first DK light curve taken through a Bessell-$I$ filter, which was shown here for comparison.}
\label{fig:DKU:lc}
\end{figure}

As concerns the two $U$-band light curves, we proceeded as in the previous cases, but, we used a linear law (the $U$-band LD is quite close to linear so there is no a real need for more complex LD laws). The best-fitting values for $k$ resulted to be $0.1168 \pm 0.0033$ and $0.1220 \pm 0.0066$, for the first and second light curve, respectively, which are not consistent with the values obtained from the fits of all the other light curves in the redder bands, see Table\,\ref{tab:fits}. Moreover, the duration of first $U$-band transit is longer (2.8 hr versus 2.4 hr) than the others, see Fig.\,\ref{fig:DKU:lc}. For these reasons, we did not use these two light curves hereinafter.

The NIR light curves were fit by the \citet{mandel:2002} transit model multiplied by a systematics detrending baseline function (BF). We extensively tested BF composed of different combination of state vectors, including star locations ($x$, $y$) and FWHM ($s_x$, $s_y$), airmass ($z$), time sequence ($t$). The final adopted BF was selected as the one with the lowest Bayesian information criterion (BIC) value, as listed below:
\begin{eqnarray}
BF(J) =c_0 + c_1 * y + c_2 * s_x + c_3 * z  \nonumber \\
BF(H) = c_0 + c_1 * x ~~~~~~~~~~~~~~~~~~~~~~~~~~~~ \nonumber\\
BF(K) = c_0 + c_1 * t \; .~~~~~~~~~~~~~~~~~~~~~~~~~~~ \nonumber
\end{eqnarray}
To account for the remaining correlated noise, we first rescaled the light curve uncertainties so that the best-fitting results have a reduced chi-square of unity. Then we inflated the rescaled uncertainties by a beta factor based on the time-averaging method \citep[e.g.][]{winn:2008}. The light-curve modelling was performed again with the new uncertainties. The best-fitting values of the parameters are reported in in Table\,\ref{tab:fits}.
Similar to our previous use of the GROND camera, the NIR light curves have stronger correlated noise than the optical light curves, and thus the measured transit depths are less precise. 
%

\subsection{Orbital period determination}
\label{sec_3.2}

We used the new optical light curves for refining the orbital ephemeris of WASP-74\,b. By modelling each light curve with {\sc jktebop}, we estimated the mid-transit times and their uncertainties employing a Monte Carlo approach. We did not use the two $U$-band and the two incomplete light curves as these gave less reliable timings \citep{gibson:2009}. We also combined the four timings measured by GROND in the optical bands into one more precise measurement. These mid-transit timings are reported in Table\,\ref{tab:tmin}.

\begin{table}
\setlength{\tabcolsep}{4pt}
\centering
\caption{Times of transit midpoint of WASP-74\,b and their residuals. The list only includes complete planetary-transit events observed by our team.}
\label{tab:tmin}
\begin{tabular}{lrrc}
\hline
~~~~Time of minimum  & ~~Cycle & O-C~~~~~~  & Telescope  \\
~~BJD(TDB)$-2400000$ & ~~no.~  & (BJD)~~~~~~ &            \\
\hline \\[-8pt]%
$57173.87170 \pm 0.00017$ &   0~~ & -0.00027528 & DK\,1.54\,m  \\
$57188.83671 \pm 0.00022$ &   7~~ &  0.00052303 & DK\,1.54\,m  \\
$57203.80033 \pm 0.00038$ &  14~~ & -0.00006866 & DK\,1.54\,m  \\
$57218.76466 \pm 0.00020$ &  21~~ &  0.00004965 & DK\,1.54\,m  \\
$57231.59161 \pm 0.00084$ &  27~~ &  0.00053249 & CA\,1.23\,m  \\
$57278.62132 \pm 0.00034$ &  49~~ & -0.00013710 & DK\,1.54\,m  \\
$57293.58447 \pm 0.00041$ &  56~~ & -0.00119879 & DK\,1.54\,m  \\
$57575.76799 \pm 0.00016$ & 188~~ &  0.00004364 & MPG\,2.2\,m  \\
\hline \end{tabular} \end{table}
We then chose the reference epoch to be that related to our first transit observation and fitted all mid-transit timings with a straight line, obtaining the following orbital ephemeris:
\begin{equation}
T_{0} = \,$BJD(TDB)$\,  2\,457\,173.87198(18) + 2.13774453(77)\,E , \nonumber
\end{equation}
where $E$ is the number of orbital cycles after the reference epoch, and the quantities in brackets denote the uncertainty in the final digit of the preceding number. We found that the orbital period is 0.475\,s (2.5$\sigma$) smaller than the value of 2.137750(1) reported by \citet{hellier:2015}, see Table~\ref{tab:finalparameters}.
The fit has $\chi_{\nu}^{2}= 2.49$, indicating that the linear ephemeris is not a very good match to the observations (the above uncertainties have been increased to account for this). However it is hard to interpret this as an indication of a transit time variation as our timings comprise only eight epochs. Many more transit observations are needed to get a clear picture. 
\begin{table*}
\caption{Photometric properties of the WASP-74 system derived by fitting the light curves.
The final parameters, given in bold, are the weighted means of the results for the individual datasets. Results from the discovery paper are included at the base of the table for comparison.}
\label{tab:fits}
\centering
\begin{tabular}{lccccc}
\hline
Telescope & date & filter & $r_{\mathrm{A}}+r_{\mathrm{b}}$ & $r_{\mathrm{b}}/r_{\mathrm{A}}$ & $i^{\circ}$ \\ 
\hline
%
DK 1.54\,m  & 2015 05 31 & Bessell $I$ & $0.2378 \pm 0.0054$ & $0.09110 \pm 0.00137$ & $79.19 \pm 0.40$ \\ %
DK 1.54\,m  & 2015 06 15 & Bessell $I$ & $0.2251 \pm 0.0108$ & $0.09243 \pm 0.00178$ & $80.01 \pm 0.78$ \\ %
DK 1.54\,m  & 2015 06 30 & Bessell $I$ & $0.2267 \pm 0.0076$ & $0.09113 \pm 0.00131$ & $79.88 \pm 0.57$ \\ %
DK 1.54\,m  & 2015 07 15 & Bessell $I$ & $0.2294 \pm 0.0068$ & $0.08868 \pm 0.00171$ & $79.58 \pm 0.50$ \\ %
DK 1.54\,m  & 2015 09 13 & Bessell $I$ & $0.2206 \pm 0.0132$ & $0.08696 \pm 0.00203$ & $80.19 \pm 0.99$ \\ %
DK 1.54\,m  & 2015 09 28 & Bessell $I$ & $0.2258 \pm 0.0138$ & $0.08516 \pm 0.00366$ & $79.66 \pm 1.04$ \\ [2pt] %
%
%
CA\,1.23\,m     & 2015 07 27 & Cousins $I$ & $0.2422 \pm 0.0159$ & $0.09424 \pm 0.00665$ & $78.95 \pm 1.39$ \\ %
CA\,1.23\,m     & 2017 09 08 & Johnson $B$ & $0.2019 \pm 0.0226$ & $0.10036 \pm 0.00500$ & $81.66 \pm 1.62$ \\ [2pt] %
MPG\,2.2\,m & 2016 07 06 & Sloan $g^{\prime}$ & $0.2293 \pm 0.0133$ & $0.08672 \pm 0.00347$ & $79.38 \pm 0.93$ \\ %
MPG\,2.2\,m & 2016 07 06 & Sloan $r^{\prime}$ & $0.2230 \pm 0.0171$ & $0.09182 \pm 0.00308$ & $80.02 \pm 1.03$ \\ %
MPG\,2.2\,m & 2016 07 06 & Sloan $i^{\prime}$ & $0.2251 \pm 0.0189$ & $0.09057 \pm 0.00475$ & $80.04 \pm 1.43$ \\ %
MPG\,2.2\,m & 2016 07 06 & Sloan $z^{\prime}$ & $0.2269 \pm 0.0144$ & $0.08926 \pm 0.00375$ & $79.77 \pm 1.05$ \\ [2pt] %
MPG\,2.2\,m & 2016 07 06 & $J$ &                $0.2052 \pm 0.0371$ & $0.08580 \pm 0.01098$ & $82.22 \pm 2.18$ \\ %
MPG\,2.2\,m & 2016 07 06 & $H$ &                $0.1995 \pm 0.0318$ & $0.09121 \pm 0.00868$ & $82.11 \pm 2.32$ \\ %
MPG\,2.2\,m & 2016 07 06 & $K$ &                $0.2123 \pm 0.0427$ & $0.08948 \pm 0.01398$ & $80.32 \pm 2.90$ \\ %
%
%
\hline
{\bf Final results} & &  &  $\mathbf{0.2297 \pm 0.0029 }$ & $\mathbf{0.09034 \pm 0.00063}$ & $\mathbf{79.86 \pm 0.21}$ \\
\hline
\citet{hellier:2015} & & & -- & $0.09803 \pm 0.00071$ & $79.81 \pm 0.24$  \\
\hline
\end{tabular}
\end{table*}

\section{Physical properties}
\label{sec:physical-properties}

We redetermined the physical properties of the WASP-74 planetary system following the \emph{Homogeneous Studies} approach (see \citealt{southworth:2012} and references therein). In brief, we combined together the measured parameters from the light curves ($r_{\rm A}+r_{\rm b}$, $k$, $i$, $P$), the new spectroscopic data ($T_{\mathrm{eff}}$ and [Fe/H]; Sect.\,\ref{sec_2.2}), and constraints on the properties of the host star coming from the evolutionary modelling of the parent star (we used five different sets of theoretical models). We also made use of the velocity amplitude of the RV variation, $K_{\mathrm{A}}=114.1 \pm 1.4$\,m\,s$^{-1}$ and fixed the eccentricity to zero \citep{hellier:2015}.

\begin{figure}
\centering
\includegraphics[width=\columnwidth]{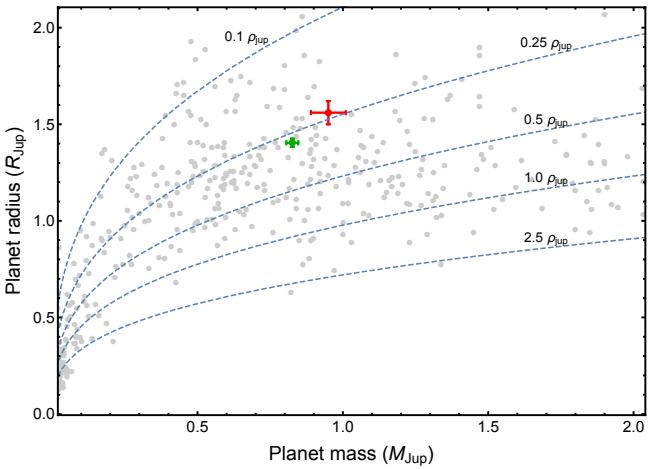}
\caption{The masses and radii of the known transiting extrasolar planets. The plot is restricted to exoplanets with masses below $2\,M_{\rm Jup}$ and radius below $2\,R_{\rm Jup}$. Grey points denote values taken from TEPCat. Their error bars have been suppressed for clarity. The position of WASP-74\,b is shown in red \citet{hellier:2015} and green (this work). Dashed lines show where density is 2.5, 1.0, 0.5, 0.25 and 0.1 $\rho_{\rm Jup}$.}
\label{fig:mass-radius}
\end{figure}

The output quantities were estimated iteratively and based on tables of stellar parameters predicted by different theoretical models over all possible ages for the star. So, at the end, we obtained a set of physical values for each of the theoretical models, and the unweighted mean of these were considered as the final values. Two uncertainties for each of the final values were assigned: a systematic error, which takes into account the level of agreement among the values obtained using different theoretical models, and a statistical error, which is related to the usual propagation of the uncertainties of the input parameters. As $\rho_{\star}$ and $g_{\rm p}$ are directly estimated from observable quantities, they do not have any systematic error. Our final values of the physical parameters of the WASP-74 planetary system are summarised in Table\,\ref{tab:finalparameters} and are significantly more precise than those estimated by \citet{hellier:2015}. In particular, our measurements point out that both planet and parent star are slightly smaller and less massive, see Fig.~\ref{fig:mass-radius}.

\begin{table*} \centering
\caption{Physical parameters of the planetary system WASP-74 derived in this work, compared with those from \citet{hellier:2015}.
Where two error bars are given, the first refers to the statistical uncertainties, while the second to the systematic errors.
{\bf Notes:} $^{(a)}$Our estimate of the stellar age was derived from theoretical models, and that from the discovery paper was
obtained from gyrochronology. $^{(b)}$The Safronov number represents the ratio of the escape velocity to the orbital velocity of the planet and indicates the extent to which the planet scatters other bodies. $^{c}$Our measurement of the time of mid-transit is given in BJD\,(TDB), while that from \citet{hellier:2015} is in HJD.}
\label{tab:finalparameters}
\begin{tabular}{l c c c c} \hline
Quantity & Symbol & Unit & This work & \citet{hellier:2015}\\
\hline  \\[-6pt]
\multicolumn{1}{l}{\textbf{Stellar parameters}} \\
Effective temperature & $T_{\rm eff}$ & K & $5984 \pm 57$ & $5990 \pm 110$ \\
Metallicity & [Fe/H] & dex & $+0.34 \pm 0.02$ & $+0.39 \pm 0.13$ \\
Projected rotational velocity & $v \sin{i_{\star}}$ & km\,s$^{-1}$ & $6.03 \pm 0.19$ & $4.1 \pm 0.8$ \\
Mass   \dotfill & $M_{\star}$ & \Msun & $1.191 \pm 0.023 \pm 0.030$ & $1.48 \pm 0.12$ \\
Radius \dotfill & $R_{\star}$ & \Rsun & $1.536 \pm 0.022 \pm 0.013$ & $1.64 \pm 0.05$ \\
Surface gravity  \dotfill     & $\log g_{\star}$& cgs  & $4.141 \pm 0.011 \pm 0.004$ & $4.180 \pm 0.018$ \\
Density \dotfill & $\rho_{\star}$ & \psun & $0.329 \pm 0.012$ & $0.338 \pm 0.018$ \\
Age$^{a}$ \dotfill & $\tau$ & Gyr & $4.2_{-0.4\,-2.0}^{+0.4\,+1.6}$   & $2.0^{+1.6}_{-1.0}$   \\
\hline \\[-6pt]%
\multicolumn{1}{l}{\textbf{Planetary parameters}} \\
Mass    \dotfill & $M_{\rm p}$ & \Mjup & $0.826 \pm 0.015 \pm 0.014$ & $0.95 \pm 0.06$ \\
Radius  \dotfill & $R_{\rm p}$ & \Rjup & $1.404 \pm 0.018 \pm 0.012$ & $1.56 \pm 0.06$ \\
Surface gravity  \dotfill & $g_{\rm p}$ & \mss & $10.38 \pm 0.26$ & $8.91 \pm 0.41$    \\
Density \dotfill & $\rho_{\rm p}$ & \pjup & $0.2788 \pm 0.0098 \pm 0.0023$ & $0.25 \pm 0.02$ \\[2pt]
Equilibrium temperature \dotfill & \Teq  & K  & $1926 \pm  21$ & $1910 \pm 40$~~~~         \\
Safronov number$^{b}$   \dotfill & \safronov\ & & $0.03396 \pm 0.00060 \pm 0.00029$ & -- \\
\hline \\[-6pt]%
\multicolumn{1}{l}{\textbf{Orbital parameters}} \\
Time of mid-transit \dotfill & $T_{0}$  & BJD(TDB) / HJD$^{c}$ & $2\,457\,173.87198 \pm 0.00018 $ & $2\,456\,506.8918 \pm 0.0002 $ \\ %
Period \dotfill & $P_{\mathrm{orb}}$ & days  & $2.1377445 \pm 0.0000018 $ & $2.137750 \pm 0.000001 $ \\ %
Semi-major axis   \dotfill   & $a$ & au    & $0.03443 \pm 0.00022 \pm 0.00029$ & $0.037 \pm 0.001 $ \\
Inclination \dotfill & $i$ & degree  & $79.86 \pm 0.21$ & $79.81 \pm 0.24$ \\  [1pt] %
\hline
\end{tabular}
\end{table*}

\section{Variation of the planetary radius with wavelength}
\label{sec_5}
%
\begin{figure}
\centering
\includegraphics[width=\columnwidth]{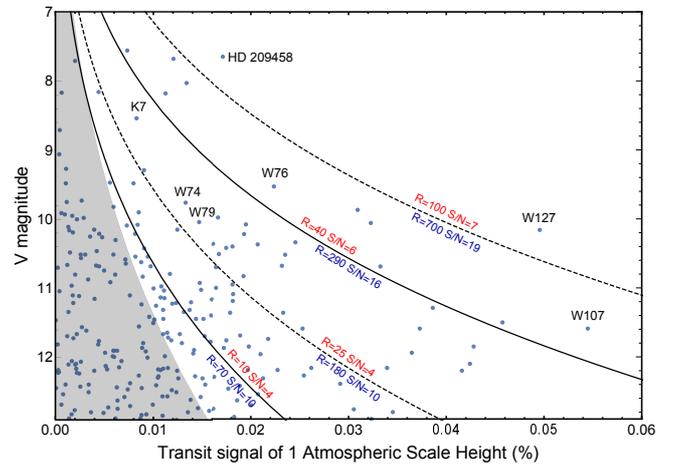}
\caption{Expected transmission spectral signal of 1 atmospheric scale height for transiting exoplanets, based on parent star $V$ magnitude. Data taken from TEPCat; error bars have been suppressed for clarity. The curves are taken from \citet{sing:2018} and indicate (approximately) constant values for S/N. The numbers in red and blue refer to HST and JWST, respectively. The gray shaded region indicate the region of the diagram in which exoplanetary atmospheres are unobservable due to low S/N. The positions of some giant exoplanets are marked.}
\label{fig:transmission}
\end{figure}

Hot Jupiters can show transmission spectra with absorption features at particular wavelengths, like sodium at $\approx590$\,nm, potassium at $\approx770$\,nm, and water vapour at $\approx950$\,nm. However, some of them can be mostly flat, suggesting a planetary atmosphere with high-altitude clouds with large-sized ($\sim \mu$m) particles, or, in other cases, it is possible to measure a slope due to the presence of small particles ($<1\, \mu$m) clouds, often referred to as hazes. There are also planets showing both clouds and hazes.

As WASP-74\,b is an inflated planet, we estimated the expected signal of its transmission spectrum, by calculating the contrast in area between the annular region of the atmosphere observed during transit and that of the star. This can be obtained from the characteristic length scale of the atmosphere, which is given by the pressure scale height
\begin{equation}
H = \frac{k_{\rm B}T_{\rm eq}}{\mu_{\mathrm{m}}\,g_{\mathrm{p}}},
\end{equation}
where $k_{\rm B}$ is Boltzmann's constant and $\mu_{\mathrm{m}}$ is the mean molecular weight, which for an atmosphere dominated by H$_2$ and He is approximately $2.3$\,amu \citep{lecavelier:2008}. Therefore, the absorption signal of the annular area of one atmospheric scale height $H$ during transit can be approximated as \citep{bento:2014}
\begin{equation}
A = \frac{2\,R_{\rm p}\,H}{R_{\star}^2}.
\end{equation}
In Fig.~\ref{fig:transmission} we plotted $A$ against the magnitude of the host stars to compare the relative signal-to-noise of WASP-74\,b with those of the other transiting exoplanets (data taken from TEPCat). The expected atmospheric transit signal for WASP-74\,b, in terms of expected S/N, is comparable to that of WASP-79\,b and KELT-7\,b, a nominal and a back-up target, respectively, for the JWST Early Release Science (ERS) program \citep{bean:2018}. WASP-74\,b is, therefore, an interesting giant planet for investigating the atmospheric composition via transmission-spectroscopy.

The transmission spectrum of WASP-74\,b was recently studied with the HST/WFC3 at NIR wavelengths ($1.1-1.7 \, \mu$m) by \citet{tsiaras:2018}. Joining this set of data with those coming from the analysis of our light curves, acquired through different passbands, we attempted to construct the transmission spectrum of WASP-74\,b over a much wider wavelength range.

\begin{table}
\centering
\setlength{\tabcolsep}{4pt}
\caption{Values of $k$ for each of the passbands. {\bf Note:} the value related to the DK telescope was obtained by simultaneously fitting the six light curves.}
\label{tab:transmission}
\begin{tabular}{lcccccccccccc} \hline
Telescope & Filter/bin & $\lambda_{\rm cen}$ (nm) & FWHM (nm) & $k$  \\
\hline
MPG 2.2\,m & Sloan $g^{\prime}$ &  459 & 149 & $0.09218 \pm 0.00110$ \\
MPG 2.2\,m & Sloan $r^{\prime}$ &  622 & 169 & $0.09564 \pm 0.00183$ \\
MPG 2.2\,m & Sloan $i^{\prime}$ &  764 & 100 & $0.09499 \pm 0.00094$ \\
CA 1.23\,m & Cousins $I$        &  787 & 110 & $0.09281 \pm 0.00100$ \\
DK 1.54\,m & Bessell $I$        &  798 & 122 & $0.09320 \pm 0.00022$ \\
MPG 2.2\,m & Sloan $z^{\prime}$ &  899 & 127 & $0.09352 \pm 0.00130$ \\[2 pt]
MPG 2.2\,m & $J$                & 1240 & 237 & $0.09014 \pm 0.00157$ \\
MPG 2.2\,m & $H$                & 1647 & 270 & $0.09300 \pm 0.00162$ \\
MPG 2.2\,m & $K$                & 2170 & 303 & $0.09062 \pm 0.00189$ \\[2 pt]
HST/WFC3 & bin~1~\,\,  & 1126 & ~~20 & $0.09094 \pm 0.00028$ \\
HST/WFC3 & bin~2~\,\,  & 1156 & ~~20 & $0.09097 \pm 0.00024$ \\
HST/WFC3 & bin~3~\,\,  & 1184 & ~~20 & $0.09068 \pm 0.00026$ \\
HST/WFC3 & bin~4~\,\,  & 1212 & ~~20 & $0.09096 \pm 0.00031$ \\
HST/WFC3 & bin~5~\,\,  & 1238 & ~~20 & $0.09119 \pm 0.00023$ \\
HST/WFC3 & bin~6~\,\,  & 1266 & ~~20 & $0.09088 \pm 0.00021$ \\
HST/WFC3 & bin~7~\,\,  & 1292 & ~~20 & $0.09124 \pm 0.00022$ \\
HST/WFC3 & bin~8~\,\,  & 1319 & ~~20 & $0.09147 \pm 0.00024$ \\
HST/WFC3 & bin~9~\,\,  & 1345 & ~~20 & $0.09080 \pm 0.00025$ \\
HST/WFC3 & bin~10 & 1372 & ~~20 & $0.09157 \pm 0.00021$ \\
HST/WFC3 & bin~11 & 1400 & ~~20 & $0.09127 \pm 0.00024$ \\
HST/WFC3 & bin~12 & 1428 & ~~20 & $0.09117 \pm 0.00027$ \\
HST/WFC3 & bin~13 & 1457 & ~~20 & $0.09156 \pm 0.00020$ \\
HST/WFC3 & bin~14 & 1487 & ~~20 & $0.09115 \pm 0.00027$ \\
HST/WFC3 & bin~15 & 1518 & ~~20 & $0.09071 \pm 0.00024$ \\
HST/WFC3 & bin~16 & 1551 & ~~20 & $0.09084 \pm 0.00020$ \\
HST/WFC3 & bin~17 & 1586 & ~~20 & $0.09122 \pm 0.00027$ \\
HST/WFC3 & bin~18 & 1625 & ~~20 & $0.09071 \pm 0.00023$ \\
\hline \end{tabular} %
\end{table}

\begin{figure}
\centering
\includegraphics[width=\columnwidth]{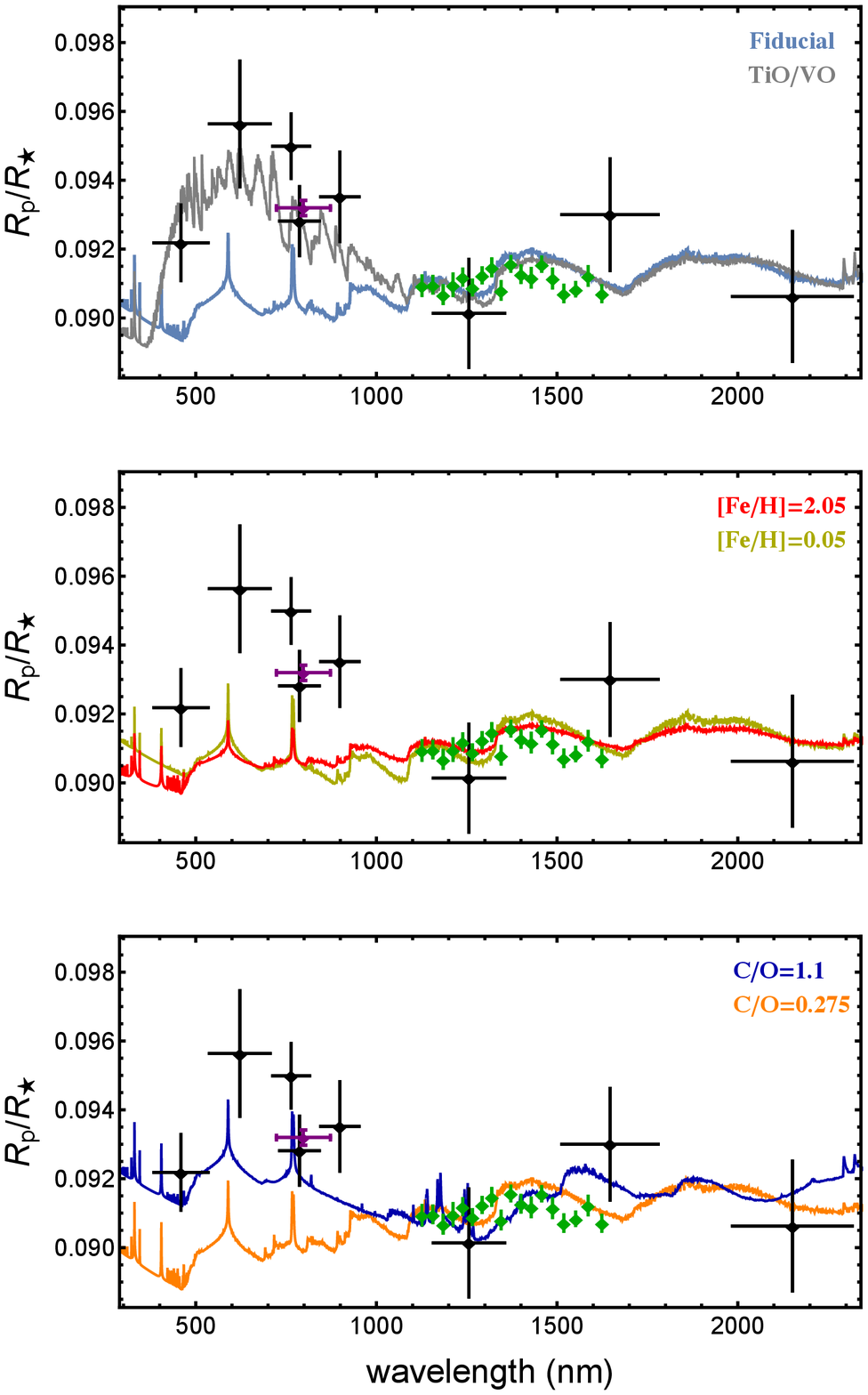}
\caption{Measurements of the planetary radius compared to the theoretical models from the {\it petitCODE}. The black points show the measured $R_{\mathrm{p}}/R_{\star}$ from this work, while the green points are from \citet{tsiaras:2018}. We marked the DK I-band value with a purple point to highlight its high precision. The vertical error bars represent the relative uncertainty and the horizontal error bars represent the FWHM of the corresponding passband. Clear models are represented separately in each plot (see Table~\ref{tab:models}): fiducial (light blue), TiO/VO (gray), metallicity 10 times more (red), metallicity 10 times less (yellow), twice solar C/O ratio (dark blue), half solar C/O ratio (orange).}
\label{fig:radiusvariation_1}
\end{figure}

\begin{figure}
\centering
\includegraphics[width=\columnwidth]{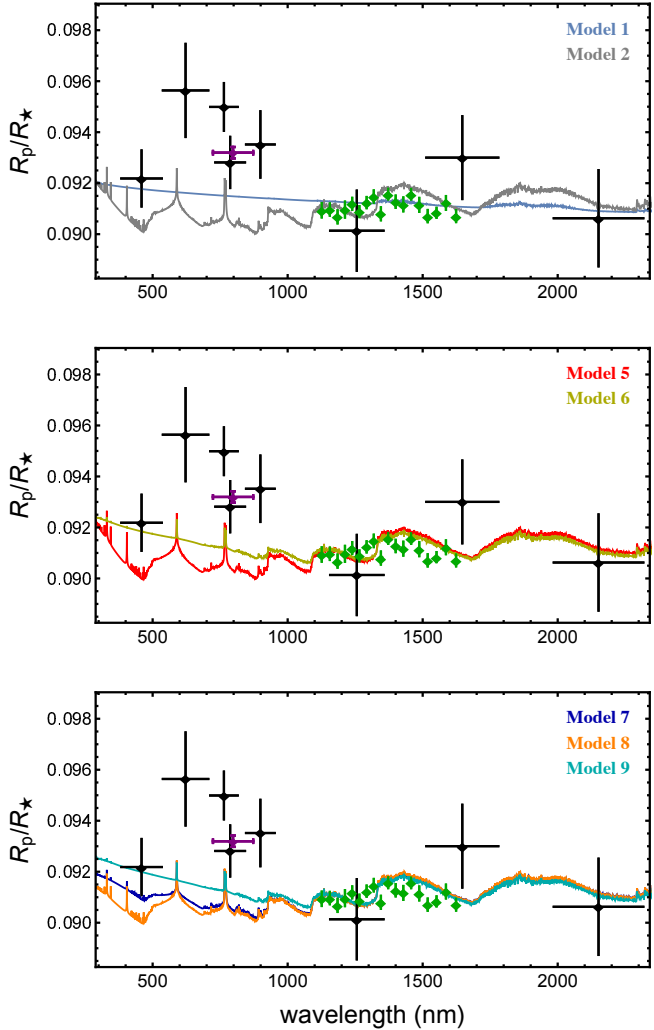}
\caption{Measurements of the planetary radius compared to the theoretical models from the {\it petitCODE}. The black points show the measured $R_{\mathrm{p}}/R_{\star}$ from this work, while the green points are from \citet{tsiaras:2018}. We marked the DK I-band value with a purple point to highlight its high precision. The vertical error bars represent the relative uncertainty and the horizontal error bars represent the FWHM of the corresponding passband. Cloud models from \citet{molliere:2017} are represented separately in each plot (see Table~\ref{tab:models}): model~1 (light blue), model~2 (gray), model~5 (red), model~6 (yellow), model~7 (dark blue), model~8 (orange), model~9 (cyan).}
\label{fig:radiusvariation_2}
\end{figure}

Following the general approach, which we also adopted in our previous studies (e.g.\ \citealp{mancini:2014,southworth:2015,tregloan:2018}), we made a new fit of the light curves to estimate the ratio of the radii, $k$, but fixing the other photometric parameters to their best values (Tables \ref{tab:fits} and \ref{tab:finalparameters}). The uncertainties of $k$ were estimated by performing 20\,000 Monte Carlo simulations. With this procedure, it is possible to obtain values of $k$ whose errorbars do not include common sources of uncertainty. In the case of the six complete transits observed with the Danish telescope, the light curves were simultaneously fitted using the {\sc jktebop} code again. We excluded from the analysis the $B$-band dataset, because the monitoring of this transit event was incomplete and the two $U$-band datasets because of their low quality. Finally, we refitted the light curves from \citet{tsiaras:2018} using the geometric parameters obtained in this work. This step is necessary to avoid possible systematic offsets between our and their data. The resulting transmission spectrum shows a large absorption in the $r$ band that, even though it is not significant ($<3\sigma$), could be an indication of TiO absorption. The values of $k$, which were obtained from the new fits, are reported in Table\,\ref{tab:transmission} and plotted in Fig.\,\ref{fig:radiusvariation_1} and Fig.\,\ref{fig:radiusvariation_2}, in which the vertical and the horizontal bars represent the relative errors in the measurements and the full widths at half maximum transmission of the passbands, respectively. We compared our results with several synthetic spectra, which were obtained with the {\it petitCODE}, a self-consistent 1D code for calculating atmospheric structures and spectra \citep{molliere:2015,molliere:2017}.

Making use of the method described in Sect~4.1 of \citet{molliere:2017}, we calculated various self-consistent clear and cloudy atmospheric structures and transmission spectra. We adopted the system parameters listed in Table~\ref{tab:finalparameters} and assumed a fiducial atmosphere with solar C/O\footnote{A gas with solar metallicity has a carbon-to-oxygen (C/O) ratio equal to 0.54 \citep{asplund:2009}.} and atmospheric enrichment of [Fe/H] $=$ 1.05 for WASP-74\,b. Models of varying metallicity ($0.1\, \times$ or $10\, \times$ that of the fiducial case), C/O ratio (solar, twice solar, half solar), with or without TiO/VO opacities, and models of various cloud setups were investigated. They are summarised in Table~\ref{tab:models}, where the cloud model number corresponds to Table~2 in \citet{molliere:2017}.

\begin{table}
\centering
\caption{Best-fitting statistics for the theoretical transmission spectra. The theoretical transmission spectra are split into two groups: cloudless clear spectra and various cloudy models with Na$_2$S and KCl and silicate clouds (see Table~2 in \citealp{molliere:2017}). Only model 8 contains iron. The best-fitting spectrum is highlighted in bold.}
\label{tab:models}
\begin{tabular}{lc} \hline
Model spectra & Best fit ($\chi_{\nu}^{2}$) \\
\hline  \\[-6pt]
\multicolumn{1}{l}{\textbf{Clear models}} \\ [2pt]
Fiducial model & 11.05\\
TiO/VO & ~~{\bf 3.72} \\
$\mathrm{[Fe/H]} = 2.05$ & ~~8.47 \\
$\mathrm{[Fe/H]} = 0.05$ & 10.51 \\
Twice solar C/O ratio & ~~8.92 \\
Half solar C/O ratio & 12.53 \\ [2pt]
\multicolumn{1}{l}{\textbf{Cloud models} (see Table~2 in \citealt{molliere:2017})} \\ [2pt]
Model 1 & ~~4.42 \\
Model 2 & 10.23 \\
Model 5 & 10.47 \\
Model 6 & ~~6.59 \\
Model 7 & ~~9.51 \\
Model 8 & 10.26 \\
Model 9 & ~~6.47 \\
\hline
\end{tabular} %
\end{table}

Following \citet{tregloan:2018}, for each model we calculated the respective $\chi^2$ using an MCMC routine to fit the models to radius ratio $k=R_{\mathrm{p}}/R_{\star}$. We first calculated the passband averages for the models and compared them to the $k$ measurements in those bands. We then fitted for a single parameter, a y-axis (radial) offset to shift the model up or down to get a best fit. Finally, the $\chi^2$ was obtained by summing over all the $k$ measurements\footnote{The number of degrees of freedom used in this work was 26.}.  The reduced chi squared, $\chi_{\nu}^2$, from fitting the 13 theoretical transmission spectra to $k$ measurements, are reported in Table~\ref{tab:models}.


Scenarios with the fiducial model and the corresponding models obtained varying metallicity and C/O ratio appear to be ruled out, as well as those with very cloudy atmospheres. The best model resulted to be that with absorption from Titanium and Vanadium oxides. However, the quality and the wavelength coverage of the data do not allow us to obtain a good constraint of the chemical composition of the atmosphere of WASP-74\,b. 
So, our results are not conclusive, but represent a first probe of the atmosphere of WASP-74\,b, and could be used together with new data.

\section{Summary}
\label{sec_6}
%
We reported new broad-band photometric light curves of eleven transit events in the WASP-74 exoplanetary system, which is composed of a F star and an inflated hot Jupiter. Most of the transits were observed with the Danish\,1.5\,m telescope through a Bessell-$I$ filter. Another transit was observed with the GROND instrument, mounted on the MPG 2.2\,m telescope, which allows observations in four optical and three NIR bands simultaneously. Finally two transits were observed in Cousins-$I$ and Johnson-$B$ filters with the CA 1.23\,m telescope, and other two with the Danish\,1.5\,m telescope through a Bessell-$U$ filter. In all the cases the defocussing technique was performed. New spectra of the star were acquired with Feros at the MPG 2.2\,m telescope and analysed with {\sc zaspe}, updating the stellar atmospheric parameters. We modelled the light curves using the {\sc jktebop} code and reviewed the orbital ephemeris and the main physical parameters of the WASP-74 planetary system. Our estimations are more precise than those reported by \citet{hellier:2015}, pointing towards smaller and less massive planet and star (see Table~\ref{tab:finalparameters} and Figure~\ref{fig:mass-radius}). In particular, we obtained $R_{\rm p}=1.404 \pm 0.022 \, R_{\rm Jup}$ versus $1.56 \pm 0.06 \,R_{\rm Jup}$; $M_{\rm p}=0.826 \pm 0.021 \, M_{\rm Jup}$ versus $0.95 \pm 0.06 \,M_{\rm Jup}$; $R_{\star}=1.536 \pm 0.026 \, R_{\sun}$ versus $1.64 \pm 0.05 \,R_{\sun}$; $M_{\star}=1.191 \pm 0.038 \, M_{\sun}$ versus $1.48 \pm 0.12 \,M_{\sun}$.

We used these new photometric multi-band observations of WASP-74\,b transits to obtain wavelength-dependent measurements of $k$, that is the ratio between the planetary radius and the stellar radius, in the optical window ($384-972$\,nm) and in some NIR windows. These measurements were used in combination with HST data \citep{tsiaras:2018} to assemble an optical-to-NIR transmission spectrum of the planet in order to probe the terminator of the atmosphere of WASP-74\,b. We measured a $r$-band radius ratio higher than in the other bands, but with a non-significant level of confidence. The observed transmission spectrum was compared with theoretical expectations coming from different models for the chemistry of the planetary atmosphere using the {\it petitCODE} (see Fig.\,\ref{fig:radiusvariation_1} and Fig.\,\ref{fig:radiusvariation_2}). The comparison of $k$ from each passband to the individual theoretical atmospheres indicates that the model that gives the best match to the data is the one with TiO and VO opacity.
More accurate data, especially at the optical wavelengths, are required to confirm this indication and, therefore, obtain a clearer picture of the atmospheric composition of WASP-74\,b.

\section*{Acknowledgements}

This paper is based on observations collected with the MPG 2.2\,m telescope, with the Danish 1.54\,m telescope and with the Zeiss 1.23\,m telescope. The first two telescopes are both located at the ESO Observatory in La Silla, Chile, while the latter is at the Centro Astron\'{o}mico Hispano Alem\'{a}n (CAHA) in Calar Alto, Spain. Operation of the MPG 2.2\,m telescope is jointly performed by the Max Planck Gesellschaft and the European Southern Observatory. Operation of the Danish 1.54\,m telescope in 2015 was financed by a grant to U.G.J.\ from the Danish Natural Science Research Council (FNU).
Operations at the Calar Alto telescopes are jointly performed by the Max Planck Institute for Astronomy (MPIA) and the Instituto de Astrof\'{i}sica de Andaluc\'{i}a (CSIC). GROND was built by the high-energy group of MPE in collaboration with the LSW Tautenburg and ESO, and is operated as a PI-instrument at the MPG 2.2\,m telescope. The reduced light curves presented in this work will be made available at the CDS (http://cdsweb.u-strasbg.fr/). L.M.\ acknowledges support from the Italian Minister of Instruction, University and Research (MIUR) through FFABR 2017 fund. L.M.\ acknowledges support from the University of Rome Tor Vergata through ``Mission: Sustainability 2016'' fund. G.C.\ acknowledges the support by the National Natural Science Foundation of China (Grant No. 11503088, 11573073, 11573075) and the project ``Technology of Space Telescope Detecting Exoplanet and Life'' from National Defense Science and Engineering Bureau civil spaceflight advanced research project (D030201). I.J.\ acknowledges the Austrian Forschungsf\"{o}rderungsgesellschaft FFG project ``RASEN'' P847963. T.C.H.\ acknowledges financial support from KASI. D.M.B.\ acknowledges the support of the NYU Abu Dhabi Research Enhancement Fund under grant RE124. M.K.\ acknowledges the support of the International Postdoctoral Fellow of Independent Research Fund Denmark. CITEUC is funded by National Funds through FCT -- Foundation for Science and Technology (project: UID/ Multi/00611/2013) and FEDER -- European Regional Development Fund through COMPETE 2020 -- Operational Programme Competitiveness and Internationalisation (project: POCI-01-0145-FEDER-006922). We thank Fei Yan for useful discussion about the WASP-74\,b transmission spectrum.
The following internet-based resources were used in research for this paper: the ESO Digitized Sky Survey; the NASA Astrophysics Data System; the SIMBAD data base operated at CDS, Strasbourg, France; and the arXiv scientific paper preprint service operated by Cornell University.



\begin{thebibliography}{99}
%
\bibitem[Alexoudi et al.(2018)]{alexoudi:2018} %
Alexoudi X. et al., 2018, \aap, in press, arXiv:1810.02172
%
\bibitem[Asplund et al.(2009)]{asplund:2009} %
Asplund M., Grevesse N., Sauval A.~J., Scott P., 2009, \araa, 47, 481
%
\bibitem[Bean et al.(2018)]{bean:2018} %
Bean J.~L. et al., 2018, PASP, 130, 114402
%
\bibitem[Bento et al.(2014)]{bento:2014} %
Bento J. et al., 2014, \mnras, 437, 1511
%
\bibitem[Brahm et al.(2017)]{brahm:2017:zaspe} 
Brahm R., Jord{\'a}n A., Hartman J., Bakos G.~\'{A}, 2017, \mnras, 467, 971
%
\bibitem[Cameron(2016)]{cameron:2016} %
Cameron A.~C., 2018, Extrasolar Planetary Transits. In: Bozza V., Mancini L. Sozzetti A. (eds), Methods of Detecting Exoplanets: 1st Advanced School on Exoplanetary Science, Astrophysics and Space Science Library, vol 428, p 89-131, Springer International Publishing Switzerland, 2016
%
\bibitem[Chen(2014)]{chen:2014} 
Chen G. et al., 2014, \aap, 563, 40 
%
\bibitem[Claret(2004)]{claret:2004} %
Claret A., 2004, \aap, 424, 919
%
\bibitem[Eastman et al.(2010)]{eastman:2010} %
Eastman J., Siverd R., Gaudi B.~S., 2010, \pasp, 122, 935
%
\bibitem[Fortney(2018)]{fortney:2018} %
Fortney J., 2018, Modeling Exoplanetary Atmospheres: An Overview. In: Bozza V., Mancini L. Sozzetti A. (eds), Astrophysics of Exoplanetary Atmospheres: 2nd Advanced School on Exoplanetary Science, Astrophysics and Space Science Library, vol 450, p 51-88, Springer International Publishing Switzerland, 2018
%
\bibitem[Gibson et al.(2009)]{gibson:2009}
Gibson N.~P. et al., 2009, \apj, 700, 1078
%
\bibitem[Gibson et al.(2018)]{gibson:2018} %
Gibson N.~P., de Mooij E.~J.~W., Evans, T.~M., Merritt S., Nikolov N., Sing D.~K., Watson C., 2018, \mnras, in press, 1810.03693
%
\bibitem[Hellier et al.(2015)]{hellier:2015} %
Hellier C. et al., 2008, \aj, 150, 18 %
%
\bibitem[Kaufer \& Pasquini(1998)]{kaufer:1998} %
Kaufer A., Pasquini L., 1998, Proc. SPIE, 3355, 844
%
\bibitem[Lecavelier des Etangs et al.(2008)]{lecavelier:2008} %
Lecavelier des Etangs A., Pont F., Vidal-Madjar A., Sing D., 2008, \aap, 481, L83
%
\bibitem[Madhusudhan(2012)]{madhusudhan:2012} %
Madhusudhan N., 2012, \apj, 758, 36 %
%
\bibitem[Mancini et al.(2013)]{mancini:2013}  
Mancini L. et al., 2013, \mnras, 436, 2 %
%
\bibitem[Mancini et al.(2014)]{mancini:2014}  
Mancini L. et al., 2014a, \aap, 562, A126 %
%
\bibitem[Mancini \& Southworth(2016)]{mancini:2016a} 
Mancini L., \& Southworth J., 2016, in {\it Twenty years of giant exoplanets}, Proceedings of the Haute Provence Observatory Colloquium,
Ed. I. Boisse, O. Demangeon, F. Bouchy, L. Arnold (published online), p. 120
%
\bibitem[Mancini et al.(2016a)]{mancini:2016a} 
Mancini L., Kemmer J., Southworth J., Bott K., Molli\'{e}re P., Ciceri S., Chen G., Henning Th., 2016, \mnras, 459, 1393
%
\bibitem[Mancini et al.(2016b)]{mancini:2016b} 
Mancini L., Giordano M., Molli\'{e}re P., Southworth J., Brahm R., Ciceri S., Henning Th., 2016, \mnras, 461, 1053
%
\bibitem[Mancini et al.(2018)]{mancini:2018} 
Mancini L. et al., 2018, \mnras, 465, 843
%
\bibitem[Mandel \& Agol(2002)]{mandel:2002} 
Mandel K., Agol E., 2002, \apj, 580, L171
%
\bibitem[Molli{\`e}re et al.(2015)]{molliere:2015} %
Molli{\`e}re P., van Boekel R., Dullemond C., Henning Th., Mordasini C., 2015, \apj, 813, 47
%
\bibitem[Molli{\`e}re et al.(2017)]{molliere:2017} %
Molli{\`e}re P., van Boekel R., Bouwman J., Henning Th., Lagage P.-O., Min M., 2017, \aap, 600, A10
%
\bibitem[Parviainen et al.(2018)]{parviainen:2018} %
Parviainen H., et al., 2018, \aap, 609, A33
%
%
\bibitem[Sedaghati et al.(2017)]{sedaghati:2017} %
Sedaghati E., Boffin H.~M.~J., Delrez L., Gillon M., Csizmadia S., Smith A.~M.~S., Rauer H., 2017, \mnras, 468, 3123
%
\bibitem[Sing et al.(2016)]{sing:2016} %
Sing D.~K. et al., 2016, Nature, 529, 59
%
\bibitem[Sing(2018)]{sing:2018} %
Sing D., 2018, Observational Techniques With Transiting Exoplanetary Atmospheres. In: Bozza V., Mancini L. Sozzetti A. (eds), Astrophysics of Exoplanetary Atmospheres: 2nd Advanced School on Exoplanetary Science, Astrophysics and Space Science Library, vol 450, p 3-48, Springer International Publishing Switzerland, 2018
%
\bibitem[Southworth et al.(2007)]{southworth:2007}
Southworth J., Bruntt H., Buzasi D.~L., 2007, \aap, 467, 1215
%
\bibitem[Southworth(2011)]{southworth:2011}
Southworth J., 2011, \mnras, 417, 2166
%
\bibitem[Southworth(2012)]{southworth:2012} 
Southworth J., 2012, \mnras, 426, 1291
%
\bibitem[Southworth(2013)]{southworth:2013} 
Southworth J., 2013, \aap, 557, 119
%
\bibitem[Southworth et al.(2014)]{southworth:2014} 
Southworth J. et al., 2014, \mnras, 444, 776 %
%
\bibitem[Southworth et al.(2015)]{southworth:2015} 
Southworth J. et al., 2015, \mnras, 447, 771 %
%
\bibitem[Southworth et al.(2017)]{southworth:2017} 
Southworth J., Mancini L., Madhusudhan N., Molli\'{e}re P., Ciceri S., Henning Th., 2017, \aj, 153, 191 %
%
\bibitem[Tregloan-Reed et al.(2018)]{tregloan:2018} %
Tregloan-Reed J., Southworth J., Mancini L., Molli\'{e}re P., Ciceri S., Bruni I., Ricci D., Ayala-Loera C., Henning Th., 2018, \mnras, 474, 5485
%
\bibitem[Tsiaras et al.(2018)]{tsiaras:2018}
Tsiaras A., et al., 2018, \aj, 155, 156
%
\bibitem[Winn et al.(2008)]{winn:2008}
Winn J.~N., et al., 2008, \apj, 683, 1076
%
\bibitem[Winn et al.(2015)]{winn:2015}
Winn J.~N., Fabrycky D~.C., 2015, \araa, 53, 409
%
\end{thebibliography}

\section*{Affiliations}
{\it
$^{1}$\,Department of Physics, University of Rome Tor Vergata, Via della Ricerca Scientifica 1, I-00133, Rome, Italy \\
$^{2}$\,Max Planck Institute for Astronomy, K\"{o}nigstuhl 17, D-69117, Heidelberg, Germany \\
$^{3}$\,INAF -- Osservatorio Astrofisico di Torino, via Osservatorio 20, I-10025, Pino Torinese, Italy \\
$^{4}$\,International Institute for Advanced Scientific Studies (IIASS), Via G. Pellegrino 19, I-84019, Vietri sul Mare (SA), Italy \\
$^{5}$\,Astrophysics Group, Keele University, Keele ST5 5BG, UK \\
$^{6}$\,Leiden Observatory, Leiden University, PO Box 9513, NL-2300 RA Leiden, Netherlands \\
$^{7}$\,Centro de Astronom\'{i}a (CITEVA), Universidad de Antofagasta, Avenida U. de Antofagasta, 02800, Antofagasta, Chile  \\
$^{8}$\,Space Research Institute, Austrian Academy of Sciences, Schmiedlstrasse 6, A-8042, Graz, Austria \\
$^{9}$\,Institut f\"{u}r Geophysik, Astrophysik und Meteorologie, Karl-Franzens-Universität, Universit\"{a}tsplatz 5, 8010 Graz, Austria \\
$^{10}$\,Institut für Astro- und Teilchenphysik, Universit\"{a}t Innsbruck, Technikerstrasse 25, A-6020 Innsbruck, Austria \\
$^{11}$\,Key Laboratory of Planetary Sciences, Purple Mountain Observatory, Chinese Academy of Sciences, Nanjing 210008, China \\	
$^{12}$\,Instituto de Astrof\'{i}sica de Canarias, V\'{i}a L\'{a}ctea s/n, E-38205 La Laguna, Tenerife, Spain \\
$^{13}$\,INAF -- Osservatorio Astronomico di Bologna, Via Ranzani 1, I-40127, Bologna, Italy \\
$^{14}$\,Department of Astronomy, Stockholm University, AlbaNova SCFAB / astronomi 106 91, Stockholm\\
$^{15}$\,Dark Cosmology Centre, Niels Bohr Institute, University of Copenhagen, Juliane Maries Vej 30, DK-2100 Copenhagen, Denmark \\
$^{16}$\,Niels Bohr Institute \& Centre for Star and Planet Formation, University of Copenhagen, {\O}ster Coldgade 5, 1350 Copenhagen K, Denmark \\
$^{17}$\,Dipartimento di Fisica ``E.R. Caianiello'', Universit\`a di Salerno, Via Giovanni Paolo II 132, 84084, Fisciano (SA), Italy \\
$^{18}$\,Istituto Nazionale di Fisica Nucleare, Sezione di Napoli, 80126 Napoli, Italy \\
$^{19}$\,New York University Abu Dhabi, Saadiyat Island, Abu Dhabi, PO Box 129188, United Arab Emirates \\
$^{20}$\,Department of Earth Sciences, Meteorological Institute, Universit\"at Hamburg, Bundesstra{\ss}e 55, D-20146 Hamburg, Germany \\
$^{21}$\,Instituto de Astrof\'{i}sica, Facultad de F\'{i}sica, Pontificia Universidad Cat\'{o}lica de Chile, Av. Vicu\~{n}a Mackenna 4860, 7820436 Macul, Santiago, Chile.
 \\
$^{22}$\,SUPA, University of St Andrews, School of Physics \& Astronomy, North Haugh, St Andrews, KY16 9SS, UK \\
$^{23}$\,European Southern Observatory, Karl-Schwarzschild-Stra{\ss}e 2, 85748 Garching bei M\"unchen, Germany \\
$^{24}$\,Physics Department and Tsinghua Centre for Astrophysics, Tsinghua University, Beijing 100084, China \\
$^{25}$\,Department of Astronomy and Space Science, Chungnam National University, Daejeon 34134, Republic of Korea \\
$^{26}$\,Korea Astronomy and Space Science Institute, Daejeon 305-348, Republic of Korea \\
$^{27}$\,Astronomisches Rechen-Institut, Zentrum f\"ur Astronomie, Universit\"at Heidelberg, M\"onchhofstra{\ss}e 12-14, 69120 Heidelberg, Germany \\
$^{28}$\,Jodrell Bank Centre for Astrophysics, University of Manchester, Oxford Road, Manchester M13 9PL, UK \\
$^{29}$\,Institute for Theoretical Astrophysics, Zentrum f\"ur Astronomie, Universit\"at Heidelberg, Albert-Ueberle-Strasse 2, 69120 Heidelberg, Germany \\
$^{30}$\,CITEUC -- Centre for Earth and Space Science Research of the University of Coimbra, Geophysical and Astronomical Observatory of the University of Coimbra, 3040-004 Coimbra, Portugal \\
$^{31}$\,Las Cumbres Observatory Global Telescope, 6740 Cortona Dr., Suite 102, Goleta, CA 93111, USA \\
$^{32}$\,Department of Physics, University of California, Santa Barbara, CA 93106-9530, USA \\
$^{33}$\,Department of Physics, Sharif University of Technology, P.\,O.\,Box 11155-9161 Tehran, Iran \\
$^{34}$\,Centre of Electronic Imaging, Department of Physical Sciences, The Open University, Milton Keynes, MK7 6AA, UK \\
$^{35}$\,Institute for Astronomy, University of Edinburgh, Royal Observatory, Edinburgh EH9 3HJ, UK \\
$^{36}$\,Stellar Astrophysics Centre (SAC), Department of Physics and Astronomy, Aarhus University, Ny Munkegade 120, DK-8000 Aarhus C, Denmark \\
$^{37}$\,DTU Space, National Space Institute, Technical University of Denmark, Elektrovej 328, DK-2800 Kgs. Lyngby, Denmark \\
$^{38}$\,Yunnan Observatories, Chinese Academy of Sciences, Kunming 650011, PR China \\
$^{39}$\,Institut d'Astrophysique et de G\'eophysique, Universit\'e de Li\`ege, 4000 Li\`ege, Belgium \\
}

\bsp 
\label{lastpage}
\end{document}